\documentclass[aps,
twocolumn,
prd,
10pt,
showpacs,
showkeys,
nofootinbib,
longbibliography,
superscriptaddress,
notitlepage]{revtex4-1}
\usepackage{dcolumn}% Align table columns on decimal point
\usepackage{bm}% bold math
\usepackage{ifpdf}
\usepackage{hyperref}
\usepackage{bm}
\usepackage{booktabs}
\usepackage{diagbox}
\usepackage{natbib}
\usepackage{amsmath} % for splitting equaitons
\usepackage{xcolor, color,graphicx, graphics}
\usepackage[spanish,english]{babel}%, portuguese
\usepackage[latin1]{inputenc}
\usepackage[OT1]{fontenc}
\usepackage{latexsym,amssymb,amsmath,amsfonts}
\usepackage{makeidx}
\usepackage{epsfig,subfigure}
\usepackage{natbib}
\usepackage{epstopdf}
\usepackage{mathrsfs}
\hypersetup{colorlinks=true, linkcolor=blue, citecolor=green}
\usepackage{enumerate}
\usepackage{xcolor}
\usepackage{multirow}
 \usepackage{array}
 \usepackage{graphicx}
\usepackage{subcaption}
\usepackage{acronym}
\usepackage{orcidlink}
\usepackage{graphicx}
\usepackage{amsmath}
\usepackage{epstopdf}
\usepackage{acronym}
\usepackage{hyperref}
\usepackage{bm}
\usepackage{newtxtext,newtxmath}
\usepackage{lastpage}
\usepackage{orcidlink}
\newacro{BH}{black hole}
\newacro{AGN}{active galactic nuclei}
\newacro{MAD}{magnetically arrested disk}
\newacro{ADAF}{Advection-Dominated Accretion Flow}

\begin{document}

%%%%%%%%%%%%%%%%%%%%%%%%%%%%%%%%%%%%%%%%%%%%%%%%%%%%%%%%%%%%%%%%%%%%%%%%%%%%%%%%%%%%%%%%%%
%%%%%%%%%%%%%%%%%%%%%%%%%%%%%%%%%%%%%%%%%%%%%%%%%%%%%%%%%%%%%%%%%%%%%%%%%%%%%%%%%%%%%%%%%%

\title[Magnetized Accretion Flow]{Magnetized Accretion Flows with Low Angular Momentum Around Supermassive Black Holes}
\author{Razieh Ranjbar \orcidlink{0000-0003-1488-4890}}
\affiliation{School of Astronomy, Institute for Research in Fundamental Sciences (IPM), P.O. Box 19395-5531, Tehran, Iran}

%%%%%%%%%%%%%%%%%%%%%%%%%%%%%%%%%%%%%%%%%%%%%%%%%%%%%%%%%%%%%%%%%%%%%%%%%%%%%%%%%%%%%%%%%%
%%%%%%%%%%%%%%%%%%%%%%%%%%%%%%%%%%%%%%%%%%%%%%%%%%%%%%%%%%%%%%%%%%%%%%%%%%%%%%%%%%%%%%%%%%

\begin{abstract}
We investigate the global structure of slowly rotating accretion flows around \acp{BH} in the presence of large-scale magnetic fields. Using a steady, height-integrated transonic model with radial mass loss, we examine how magnetic pressure, tension, and torque modify the dynamics of Bondi-like and disk-like accretion. Adopting a radial magnetic-field prescription based on results from previous numerical simulations, we compare hydrodynamical and magnetized solutions across a range of plasma-$\rm \beta$ values, from weak to strong field regimes. We find that magnetic stresses substantially alter the sonic radius and change the angular momentum and mass accretion rate reaching the \ac{BH}. For sufficiently strong fields, the inner flow becomes highly sub-Keplerian and magnetically dominated ($ \beta= P_{\mathrm{gas}}/ P_{\mathrm{mag}}<1$), and magnetic stresses can drive transitions to lower angular momentum. We also show that outflows in this magnetized accretion flow can significantly alter the structure of disk-like accretion. We further perform 2D simulations of magnetized Bondi accretion and torus accretion and find that the radial structure and magnetically dominated, sub-Keplerian character of our semi-analytical solutions are qualitatively reproduced.
\end{abstract}


\maketitle

\section{Introduction}
\ac{BH} accretion is among the most efficient mechanisms for extracting gravitational energy in astrophysics, powering systems ranging from \acp{AGN} to X-ray binaries. In these systems, matter transferred toward the central \ac{BH} typically forms an accretion disk before gradually accreting onto the \ac{BH} (e.g., \citealp{FrankKingRaine2002}). In the simplest picture, gas supplied at large radii loses angular momentum, moves inward through the gravitational potential of the central \ac{BH}, and releases part of its binding energy as radiation or kinetic energy. The structure of the accretion flow, however, depends sensitively on the thermodynamic state of the gas, the angular momentum content at the outer boundary, and the magnetic flux carried inward with the plasma.

Based on gas temperature, accretion disks are broadly classified into two categories \citep{YuanNarayan2014}: a geometrically thin, optically thick cold disk that emits approximately blackbody radiation, first described by \citet{ss73} (hereafter SSD); and a geometrically thick, hot accretion flow with low optical depth and optically thin emission, such as an \ac{ADAF} \citep{Narayan1994}. These hot flows are especially relevant for low-luminosity \acp{AGN}, quiescent X-ray binaries, and massive \acp{BH} embedded in the hot atmospheres of elliptical galaxies, where accretion proceeds at low Eddington ratios through radiatively inefficient, advection-dominated flows (see \citealp{YuanNarayan2014} for a review).

Early analytical models of radiatively inefficient accretion flows assumed a constant mass accretion rate with radius, giving the self-similar density profile $\rho \propto r^{-3/2}$ \citep{Narayan1994}. Over the past two decades, however, numerous hydrodynamical and magnetohydrodynamical simulations have shown that this assumption does not generally hold: the mass inflow rate instead decreases toward the central \ac{BH}, typically following $\dot{M}(r) \propto r^{s}$ with $0 \leq s \leq 1$, commonly interpreted as evidence of significant mass loss through winds and outflows \citep{Stone1999, YuanNarayan2014}. The resulting density profile is flatter than the classical self-similar prediction, following $\rho \propto r^{-p}$ with  $p \lesssim 1$. These theoretical and numerical results are supported observationally by low-luminosity accreting systems, including the Galactic center \ac{BH} SgrA*, the AGN NGC~3115, and several \acp{BH} in elliptical galaxies \citep{Cho2024}.

Classical accretion theory, however, treats spherically symmetric Bondi accretion \citep{Bondi1952} and disc-like \acp{ADAF} as two distinct limiting cases, without a unified framework for flows of finite angular momentum. If the accretion flow carries essentially no angular momentum, it produces quasi-spherical Bondi accretion; if it carries Keplerian angular momentum, a geometrically thin disk forms instead \citep{ss73, Dihingia2025}. In practice, accreting gas usually possesses some intermediate angular momentum, and rotational effects cannot be neglected. Motivated by this gap, several studies have examined how slow rotation modifies classical Bondi accretion \citep{Jiao2026, Ranjbar2022, NarayanFabian2011, Park2009}, while others have investigated the influence of external gravitational potentials \citep{Ranjbar2025, Ranjbar2025a}. More recently, attention has turned to outflows in Bondi-like and slowly rotating accretion flows \citep{Dongho2026, Ranjbar2023, HanPark2026}.

Alongside angular momentum, magnetic fields play a fundamental role in accretion flow dynamics. It is now widely accepted that angular momentum transport in accretion disks is primarily driven by the magnetorotational instability (MRI; \citealp{BalbusHawley1991}), which requires differential rotation to operate and introduces effective viscosity through turbulence. Accretion disks generally contain small-scale turbulent magnetic fields, with length scales much shorter than the disk scale height $ H$, generated and amplified by MHD instabilities such as the MRI \citep{BalbusHawley1998}. In weakly magnetized flows, the magnetic pressure is generally subdominant to the gas pressure ($ \beta>1$), although magnetic stresses can still provide an important mechanism for angular-momentum transport \citep{BalbusHawley1998}.

Depending on the net magnetic flux threading the flow, accretion can instead settle into one of two qualitatively different regimes. With weak net flux, the flow remains in the weak-field regime \citep{Narayan2012}. If strong magnetic flux is instead dragged inward toward the central \ac{BH}, it can accumulate until it saturates the inflow, forming a \ac{MAD} \citep{Narayan2003, McKinney2012}, from which a collimated, powerful jet can be launched \citep{Tchekhovskoy2011}. Using a semi-analytical, height-integrated \ac{ADAF} model, \citet{Li2025} showed that an inner \ac{MAD} region can form self-consistently at several tens of Schwarzschild radii from the horizon via inward advection of magnetic flux, provided the external field is not too weak ($\beta \lesssim 100$), and that the resulting jet power can exceed that of a normal \ac{ADAF} by roughly two orders of magnitude. Numerical simulations further show that the geometry of the field -- purely toroidal versus purely poloidal -- can produce marked differences in the morphology and dynamics of radiatively inefficient flows (e.g., \citealp{Igumenshchev2003}). 

Despite these advances, the global dependence of magnetized accretion on intermediate angular momentum, particularly the transition between Bondi-like and disk-like flow geometries with and without outflow, remains insufficiently explored. Most previous work on magnetized flows has focused either on \acp{ADAF} with Keplerian-like angular momentum \citep{samadi2014, samadi2017} or on classical, non-magnetized Bondi accretion. 

Recent 3D GRMHD simulations have demonstrated that magnetic fields can
substantially modify low-angular-momentum accretion. In particular,
\citet{Cho2023} investigated magnetized Bondi accretion onto non-spinning \acp{BH} and found that strong magnetic flux can suppress the mass
accretion rate relative to the Bondi value. Their simulations further showed
that, as magnetic flux accumulates near the \ac{BH}, the flow can enter a
magnetically arrested state characterized by highly non-axisymmetric
interchange motions and strong magnetic braking. Subsequent simulations
\citep{Cho2024} showed that magnetic braking can substantially reduce the
dependence of the accretion flow on the initial angular momentum of the gas.
For spinning \acp{BH}, \citet{Chan2025} found a time-averaged feedback
efficiency of $\eta \approx 0.3$, largely independent of the Bondi radius and
the initial rotation state, while \citet{ChoNarayan2025} demonstrated that
these systems can develop strongly variable accretion and jet power,
including rotational flips. More recently, \citet{Galishnikova2025} showed
that strongly magnetized accretion with low angular momentum can produce
weak jets when turbulence limits the accumulation of magnetic flux.
These simulations demonstrate that even nearly spherical magnetized accretion can depart strongly from the classical Bondi solution, highlighting the need to understand how such magnetic effects evolve as finite angular momentum is introduced.

These studies demonstrate that magnetic fields can strongly alter the dynamics
of nearly spherical, low-angular-momentum accretion flows. However, they
primarily rely on time-dependent 3D GRMHD simulations of Bondi-like accretion
and focus on magnetic flux accumulation, interchange motions, and the
resulting variability. They do not provide a steady global description of how
magnetic stresses modify the transition from Bondi-like to disk-like
accretion as the angular momentum of the inflowing gas is varied. In
particular, the effects of magnetic pressure, tension, and torque on the
critical-point structure, angular-momentum distribution, and mass accretion
rate across this sequence have not been systematically explored.

In this work, we address this gap by extending our previous hydrodynamic
study of low-angular-momentum accretion to include large-scale magnetic
fields. We develop a steady, height-integrated transonic model that accounts
for magnetic pressure, magnetic tension, magnetic torque, and radial mass
loss. A related semi-analytical model was developed by \citet{Li2025}, who studied the transport of poloidal magnetic flux from an outer \ac{ADAF} to an inner \ac{MAD} region, deriving the field configuration self-consistently from the flux-transport equations for a flow with small angular momentum. In contrast, our model covers the transition from quasi-spherical, Bondi-like to rotationally supported, disk-like accretion, adopts a prescribed power-law magnetic-field geometry motivated by numerical simulations rather than solving for flux accumulation directly, and includes mass loss through outflows. We therefore focus on how magnetic-field strength affects the critical-point structure, angular-momentum distribution, and accretion rate across a broad range of rotation, including strongly and weakly magnetized flows.

We further compare the semi-analytical solutions with 2D time-dependent GRMHD simulations of magnetized Bondi and torus accretion to test whether the radial structure and sub-Keplerian, magnetically dominated character of our semi-analytical solutions are reproduced in time-dependent simulations. Because the semi-analytical model is axisymmetric and height-integrated, these
simulations are used primarily to test the radial and angular-momentum
structure of the solutions rather than to reproduce the fully
three-dimensional turbulent state.

This paper is organized as follows. In Section~\ref{sec:2}, we describe
the model of the accretion flow and the governing MHD equations. We
also present the adopted magnetic-field prescription and the iterative
procedure used to obtain the global flow solution. In Section~\ref{sec:3},
we present the results for weakly and strongly magnetized flows and
investigate the effects of outflows. We then compare our results with
numerical simulations of magnetized Bondi and torus accretion and
discuss the implications for low-angular-momentum accretion onto
supermassive black holes. Section~\ref{sec:4} presents the conclusions
of our study.

\section{Model}\label{sec:2}
\subsection{The structure of the Gas in the magnetized accretion flow}
This section outlines the fundamental equations of our
model. The radial accretion velocity of a slowly rotating accretion flow is significantly
less than that of Bondi accretion, which results in less 
inward flux transport. The magnetic fields considered here are motivated by MHD simulations and represent magnetically dominated accretion states; however, the model does not follow the time-dependent magnetic flux accumulation process responsible for \ac{MAD} formation. We adopt the cylindrical coordinate
$(R, \phi, z)$ and consider a steady ( $\partial/\partial t = 0$)
and axisymmetric ($ \partial/\partial \phi = 0$) hot accretion flow around a Schwarzschild \ac{BH}. The gas within an \ac{ADAF}-like flow is very hot and has a positive Bernoulli parameter; thus, one may expect a strong
outflow associated with its accretion process \citep{ny94, ny95, BlandfordBegelman1999}. Below, we present the dynamical equations governing the magnetized accretion flow. We restrict our analysis to the gas-filled region inside $ R_{\rm m}$, where the accreting gas spirals inward along the magnetic field lines.

\begin{equation}\label{con.eq}
\frac{d}{dR}(4 \pi \rho R H) + \frac{d\dot{M}_{\rm w}}{dR} = 0,
\end{equation}

 Due to the high ionization state of the gas, these outflows are difficult to detect. In this work, following \citet{Knigge1999}, we assume that the mass accretion rate $\dot{M}_{\rm acc}$ is a power-law dependence of the radius, which reads as

\begin{equation}
    \dot{M}_{\rm acc} = -4\pi R H \rho v = \dot{M}_{\rm out} (\frac{R}{R_{\rm out}})^s
\end{equation}

where $s$ is the mass-loss power-law index and represents the
strength of the outflow. Also, $R_{\rm out}$ and  $\dot{M}_{\rm out}$ are the radius and
the mass accretion rate in the outer boundary, respectively \citep{Ranjbar2023}.

The following new system of equations is based on the Bondi flow arising from a slowly rotating hot atmosphere, following \cite{NarayanFabian2011}. The momentum equations in the radial direction can be expressed as
\begin{equation}
v\frac{dv}{dR} + \frac{1}{\rho} \frac{d\rm P_{\mathrm{tot}}}{dR} = (\Omega^2 - \Omega^2_{\rm K})R +g_{\rm m}
\end{equation}

where $g_{\rm m}$ is the radial magnetic force caused by the curvature
of the magnetic field line, which is

\begin{equation}
g_{\rm m} = \frac{B_{\rm R} B_{\rm z}}{4 \pi \rho H}
\end{equation}

With the assumption that the accretion flow is in a stationary state in
the vertical direction, the scale height of the accretion disk, $\rm H$, can be expressed as $H = c_{\rm s}/\Omega_{\rm K}$. Here, $c_{\rm s} = (P_{\rm tot}/ \rho )^{1/2}$ is the sound speed, and the total pressure $P_{\rm tot}$ includes
two contributions, namely of the gas, $P_{\rm g}$; the magnetic field,
$ P_{\rm mag} = B^2 / 8 \pi$. The radiation pressure is
negligible in hot accretion flows. We thus write $P_{\rm tot} = P_{\rm g} + P_{\rm mag}$. 

The effects of general relativity are approximately described with the Paczynski-Wiita potential.
\begin{equation}
\psi_{\rm PW} = - \frac{GM}{R - R_{\rm s}}    
\end{equation}
where $M$ is the mass, and $R_{\rm s} = 2GM/c^2 $ denotes the
Schwarzschild radius of the non-rotating \ac{BH}.

So, the Keplerian angular velocity can be defined

\begin{equation}
    \Omega_{\rm K} = \frac{(GM)^{1/2}}{R^{1/2}(R - R_{\rm s})}
\end{equation}

where $\Omega$ is the angular velocity of the gas on the disc mid-plane; we mainly focus on a condition of $\Omega^2 << \Omega^2_{\rm K}$, meaning the centrifugal acceleration is extremely weak compared to the gravitational one.

The momentum equations in azimuthal
directions can be expressed as

\begin{equation}
\frac{d}{dR}(4 \pi R^4 \rho \nu \Omega) = \frac{d}{dR}(2 \pi R \nu \Sigma R^2 \frac{d\Omega}{dR}) + 2\pi R T_{\rm m} - (lR)^2 \Omega \frac{d \dot{M}_{\rm w}}{dR},
\end{equation}

where $ \Sigma = 2 \rho H $ is the surface density, $\nu=\alpha c_{\rm s} H$ is the kinematic viscosity, $\alpha$ is the dimensionless viscosity parameter, and $\Omega$ is the angular velocity of the gas within the accretion flow. The parameter $l$ is a dimensionless quantity describing the specific angular momentum carried away by the outflow.

$T_m$ is the torque exerted by the global magnetic field, which transports angular momentum through the accretion flow.

\begin{equation}\label{T_{m}}
T_m = \frac{R}{H} \frac{B_{\rm z} B_{\phi}}{4 \pi}
\end{equation}

Correspondingly, we introduce  plasma gas-to-magnetic parameters,
$\beta = \frac{P_{\mathrm{gas}}}{P_{\mathrm{mag}}}$

In the scenarios considered in this paper, the accretion flow is expected to be radiatively inefficient, such that the energy equation can be expressed as

\begin{equation}\label{energy.eq}
    \rho v \frac{d\varepsilon}{dR} - c_{\rm s}^2 v \frac{d\rho}{dR} = \frac{f \alpha \rho c_{\rm s}^2 R^2}{\Omega_{\rm K}} (\frac{d\Omega}{dR})^2 - \frac{1}{2}\eta\dot{m}_{\rm w} v_K^2
\end{equation}

where the specific internal energy of the gas is given by $\varepsilon = \frac{c_{\rm s}^2}{\gamma-1}$, $\gamma$ is the ratio of specific heats, which we set to $\gamma=5/3$, and $\eta$ is a free, dimensionless parameter characterizing the energy carried away by the outflow. In the above equation, following \cite{ny94}, the advection parameter is introduced as $f = Q_{\rm adv}/Q_+$. Consequently, the cooling term takes the $(1 - f)$ factor of the heating rate, which is an excellent approximation for radiatively inefficient accretion flows (RIAFs).
We set $f = 1$ throughout this paper (full advection case).

\subsection{Global Magnetic Field Configuration}

To close the system, we prescribe the large-scale magnetic-
field geometry following the behaviour seen in numerical
simulations of magnetically arrested and SANE accretion
flows (\cite{McKinney2012, XieZdziarski2019}). We thus set the field structure based on the 3D
MHD simulations of \cite{McKinney2012}. We thus express $B_{\rm z} (R)$, defined as
the strength of $\rm B_{\rm z}(R)$ azimuthally averaged over the spirals and
voids, as

\begin{equation}
B_{\rm z}(R) = B_{\rm z0} (\frac{R}{R_{\rm 0}})^{-s_{\rm bz}}
\end{equation}

where we assume $s_{\mathrm{bz}}$ based on the results of \cite{McKinney2012}.
With the assumption that $B_{\mathrm{R}}(R,z)$ is an odd function of $\rm z$ (adopted in this work and also
in \citealp{Lubow1994}), they further found $\rm B_{\rm R}$ to be

\begin{equation}
B_{\rm R} = \frac{1}{\mathcal{P}_{\rm m}} \frac{H}{R} B_{\rm z}
\end{equation}
we set  $\mathcal{P}_{\rm m}=2$  as the fiducial value. With a given $ B_{\rm z}$, we have a relatively weaker $ B_{\rm R}$. This is consistent with MHD simulations of \ac{MAD}; see \cite{McKinney2012}

The azimuthal magnetic field component is prescribed following the formulation adopted by \cite{XieZdziarski2019}. In this prescription, the $\phi$-component $B_{\rm \phi}$ directly connects to the rotation of the accreting gas. We have
\begin{equation}
B_{\phi} = \kappa_{\rm \phi} \frac{\Omega R}{|v| + \Omega_{\rm K} R} \frac{H}{R} B_{\rm z}
\end{equation}
where $\kappa_{\rm \phi}$ is a free parameter. In this expression, the Keplerian
velocity $\Omega_{\rm K} R$ is introduced to reduce the $B_{\rm \phi}$ value in case of
slower rotation. Also, this form ensures $B_{\rm \phi}$ is always smaller
than the dominant component $B_{\rm z}$. Because $B_{\rm R} B_{\rm \phi} < 0$ , we take a
negative value of $\kappa_{\rm \phi} = -0.5$ as a fiducial value

\begin{figure*}
    \centering

    \includegraphics[width=0.35\textwidth]{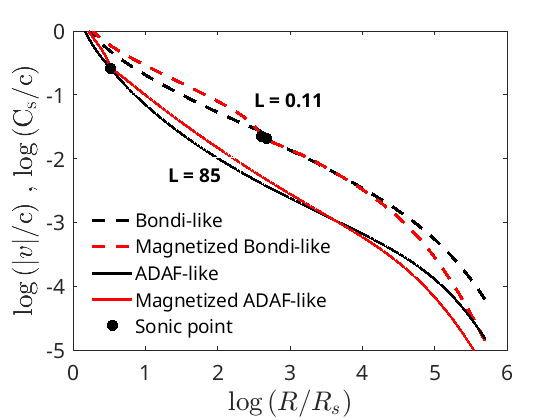}
    \includegraphics[width=0.35\textwidth]{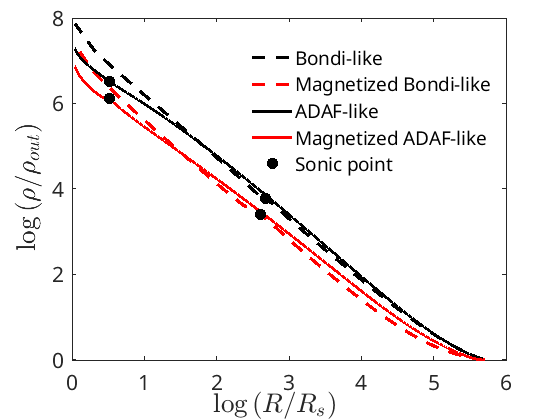}

    \medskip

    \includegraphics[width=0.35\textwidth]{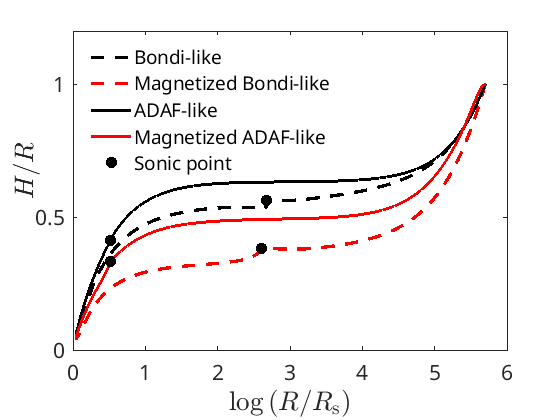}
    \includegraphics[width=0.35\textwidth]{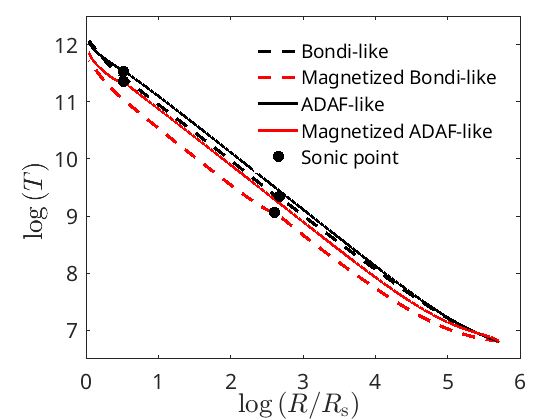}
\caption{
Comparison of the radial structure of Bondi-like and \ac{ADAF}-like accretion
solutions in the hydrodynamic and magnetized cases for
$ \alpha=0.1$, $ \beta=0.9$, and $\rm s=0$.
The four panels show the radial profiles of the radial velocity
(upper left), density (upper right), relative geometrical thickness
$\rm H/R$ (lower left), and temperature (lower right).
Black curves denote the non-magnetized solutions, while red curves
represent the corresponding magnetized solutions. Dashed and solid
lines indicate the Bondi-like ($\rm L=0.11$) and \ac{ADAF}-like ($\rm L=85$)
branches, respectively. Filled circles mark the sonic points.
The corresponding specific angular momenta swallowed by the \ac{BH} and critical radii are
$ j=0.3609$, $R_{\rm c}=467.5$ for the non-magnetized Bondi-like solution,
$ j=0.070$, $ R_{\rm c}=401.5$ for the magnetized Bondi-like solution,
$ j=1.042$, $ R_{\rm c}=3.302$ for the non-magnetized \ac{ADAF}-like solution,
and $ j=0.591$, $R_{\rm c}=3.249$ for the magnetized \ac{ADAF}-like solution.
The corresponding mass accretion rates in our units (where $ R_{\rm s} =  c = \rho_{\rm out} = 1$) are
$\dot{M}=1.93\times10^{8}$, $4.34\times10^{7}$,
$4.71\times10^{7}$, and $1.523\times10^{7}$, respectively.
}
\label{Fig2}
\end{figure*}

\subsection{Boundary conditions}

For the boundary condition, we use two boundaries: one is at the outer boundary, and the other is at the sonic point.  For the outer boundary, we assume it is around the Bondi radius, and it connects to the interstellar medium. 

We need to solve the combination of differential Equations \ref{con.eq} and \ref{energy.eq}
while simultaneously determining the mass accretion rate $\dot{M}$, the specific angular momentum at the inner edge of flow $ j$, and the critical radius $ R_{\rm c}$, all for a given outer
boundary conditions at $R =  R_{\rm out}$, i.e.,

\begin{equation}
  c_{\rm s} =  c_{\rm s,{\rm out}}, \quad \Omega = \Omega_{\rm out} 
\end{equation}

In our calculations, we adopt the no-torque condition at the
horizon and the assumption that angular momentum transport
is minimal in the supersonic part of the flow, as adopted
in \cite{NarayanFabian2011}. As a consequence, the specific angular momentum
remains constant for $R < R_{\rm c}$ and the flow has to satisfy
$d(\Omega R^2)/dR = 0$ as an additional boundary condition at $R = R_{\rm c}$.

The iterative integration proceeds as follows. We first
choose the adiabatic index $\gamma$, the viscosity parameter $\alpha$, the
outer radius $R_{\rm out}$, and the dimensionless angular momentum at the outer
boundary $L$. 
We assume a trial value for $R_{\rm c}$ and integrate the combination of Equations~(\ref{con.eq})--(\ref{energy.eq}) from $R_{\rm out}$ to $ R_{\rm c}$ using the relaxation method. When the integrated solution diverges due to an
incorrect trial value for $ R_{\rm c}$, we adjust $ R_{\rm c}$ and repeat the
relaxation procedure until the solution converges. The mass
accretion rate $\dot{M}$ and the specific angular momentum $j$ are
determined self-consistently through relaxation. Once $R_{ \rm c}$ is
determined, we apply L'Hôpital's rule to pass the critical point
and then integrate inward down to the inner boundary with a
usual integration driver.
We set the outer boundary at the Bondi radius,
$R_{\mathrm{out}} = R_{\mathrm{B}}$, the temperature at $R_{\rm{out}}$ is
$T_{\rm out} = \mu m_{\rm p} c_{\rm s,\rm out}^2/k_{\rm B} = 2.73 \times 10^6 $ $\mathrm{K}$, and the
inner boundary at the $R_{\rm in} = 1.1 \ R_s{\rm s}$. Under the assumption that the plasma $\beta$ remains locally constant, the magnetic pressure contributes an additional pressure support term, leading to an effective sound speed modified by the factor $1+1/\beta$. 

\begin{equation}
    \frac{d \ln v}{dR}\left(v^2 - c_{\rm s}^2(1+\frac{1}{\beta})\right) = \frac{(\Omega_{\rm K}^2 - \Omega^2)R}{c_{\rm s}^2} \\ 
    \end{equation}
    \begin{equation*}
     - \frac{2\gamma}{\gamma+1}(1 + \frac{1}{\beta}) (\frac{1}{R} - \frac{d\Omega_{\rm K}}{dR}) + (1 +\frac{1}{\beta}) \frac{\gamma -1}{\gamma +1} \frac{\nu R^2}{v c_{\rm s}^2} (\frac{d\Omega}{dR})^2 - \frac{g_{\rm m}}{c_{\rm s}^2} 
\end{equation*}

Since $dv/dR$ must be well-behaved across the sonic point, the regularity condition at the critical point provides two
additional boundary conditions at $ \rm R_{\rm c}$:

\begin{equation}
    v^2 - \frac{2 \gamma}{\gamma+1} (1 +\frac{1}{\beta}) c_{\rm s}^2 =0, \quad R = R_{\rm c}
\end{equation}

\begin{equation}
        \frac{(\Omega_{\rm K}^2 - \Omega^2)R}{c_{\rm s}^2} - \frac{2\gamma}{\gamma+1}(1 + \frac{1}{\beta}) (\frac{1}{R} - \frac{d\Omega_{\rm K}}{dR}) \\
\end{equation}
\begin{equation*}
    +(1 +\frac{1}{\beta}) \frac{\gamma -1}{\gamma +1} \frac{\nu R^2}{v c_{\rm s}^2} (\frac{d\Omega}{dR})^2 - \frac{g_{\rm m}}{c_{\rm s}^2} = 0 ,\quad R = R_{\rm c}
\end{equation*}

Our model accretion flow consists of a viscous subsonic region which extends from the
sonic radius $ R_{\rm c}$ out to some large outer radius $ R_{\rm out}$. Finding the solution in the viscous region requires solving a boundary value problem involving a number of differential
equations. The above equations can be combined to obtain three first-order ordinary differential equations for the three unknown radial profiles, $\Omega(r)$, $c_{\rm s}(r)$, and $v(r)$, requiring three boundary conditions.

In addition, the constants $ R_{\rm c}$ (the sonic radius) and $\rm j$ (the specific angular momentum) are eigenvalues of the problem, and their determination requires two additional boundary conditions. Physically, 
$\rm j$ represents the specific angular momentum swallowed by the \ac{BH} under the no-torque condition imposed just outside the \ac{BH} horizon \citep{Narayan1997}.

One way to reduce the standard boundary value problem is as follows \citep{Press1992, Ranjbar2023}. Here, only one boundary point, $R_{\rm out}$, is specified, while the other boundary, $ R_{\rm c}$
, is to be determined so that the system admits a solution satisfying a total of $\rm N+1$ boundary conditions.

%%%

\subsection{Model Parameter}

The model parameters are as follows. We set the \ac{BH} mass to $ M_{\rm BH}=10^8 \,  M_{\rm \odot}$, and adopt an advection parameter of $\rm f=1$ throughout, so that the flow is fully advective. The outer boundary is placed at the Bondi radius, $ R_{\rm out}=  R_{\rm B} \sim 10^{5} \, R_{\rm s}$. Two boundary conditions are imposed at $R_{\rm out}$. One is the angular momentum at the outer boundary, defined as dimensionless $L=l_{\rm out}/l_{\rm ms}$, where $l_{\rm out}$ is the specific angular momentum at the outer boundary and $l_{\rm ms}$ is the specific angular momentum at the marginally stable orbit. The other is the gas temperature, which is fixed at $T_{\rm out}=2.73\times10^6 \, \mathrm{K}$, a value characteristic of the hot atmosphere of a nearby elliptical galaxy. The inner boundary is placed at $ R_{\rm in}= 1.1 \,  R_{\rm s}$. We adopt $\alpha=0.1$ throughout this study.

Two parameters are varied systematically. The first is the plasma parameter, $\beta = P_{\rm g}/P_{\rm {mag}}$, which measures the strength of the large-scale magnetic field and is scanned over the range $10^{-3} \leq \beta \leq 5 \times 10^2$, spanning the regime from a strongly magnetically dominated flow to an essentially weakly magnetized one. The second is the mass-loss index, $\rm s$, which is varied between $0$ and $0.5$, where $\rm s = 0$ corresponds to the absence of mass loss by outflows, while larger values indicate increasing outflow intensity. $\rm l$ and $\eta$ are the angular momentum and energy parameters of the outflow, respectively, both set to $1.0$. One additional parameter is the radial distribution of the vertical magnetic field $B_{\rm z}$. We adopt a power-law profile with a slope of $s_{\rm bz}=1.1$.

Note that if the global magnetic field is relatively weak (i.e., $R_{\rm m}<  R_{\rm out}$), the system remains nearly axisymmetric at the outer boundary, with $C_{\rm m}=1$ and $B_{\rm z}=  B_{z,\rm out}$.

%%%
\section{Results}\label{sec:3}

We numerically solved the dynamical structure of the \ac{ADAF}-like and Bondi-like accretion flows in the presence of a magnetic field.  In the calculations, we set the outer boundary of the \ac{ADAF} at
$R_{\rm out} \sim  R_{\rm B} \sim 10^{5} \, R_{\rm s}$  and the corresponding outer boundary conditions, i.e., $\rho_{\rm out}$, $c_{\rm s,\rm out}$, and $\Omega_{\rm out}$, are preset as described by
the global solution in \cite{NarayanFabian2011}. We
adopt $\alpha = 0.1$, and a moderately external magnetic field with $10^{-3} \leq \beta \leq 5 \times 10^2$. 

\begin{figure*}
    \centering

   \includegraphics[width=0.4\textwidth]{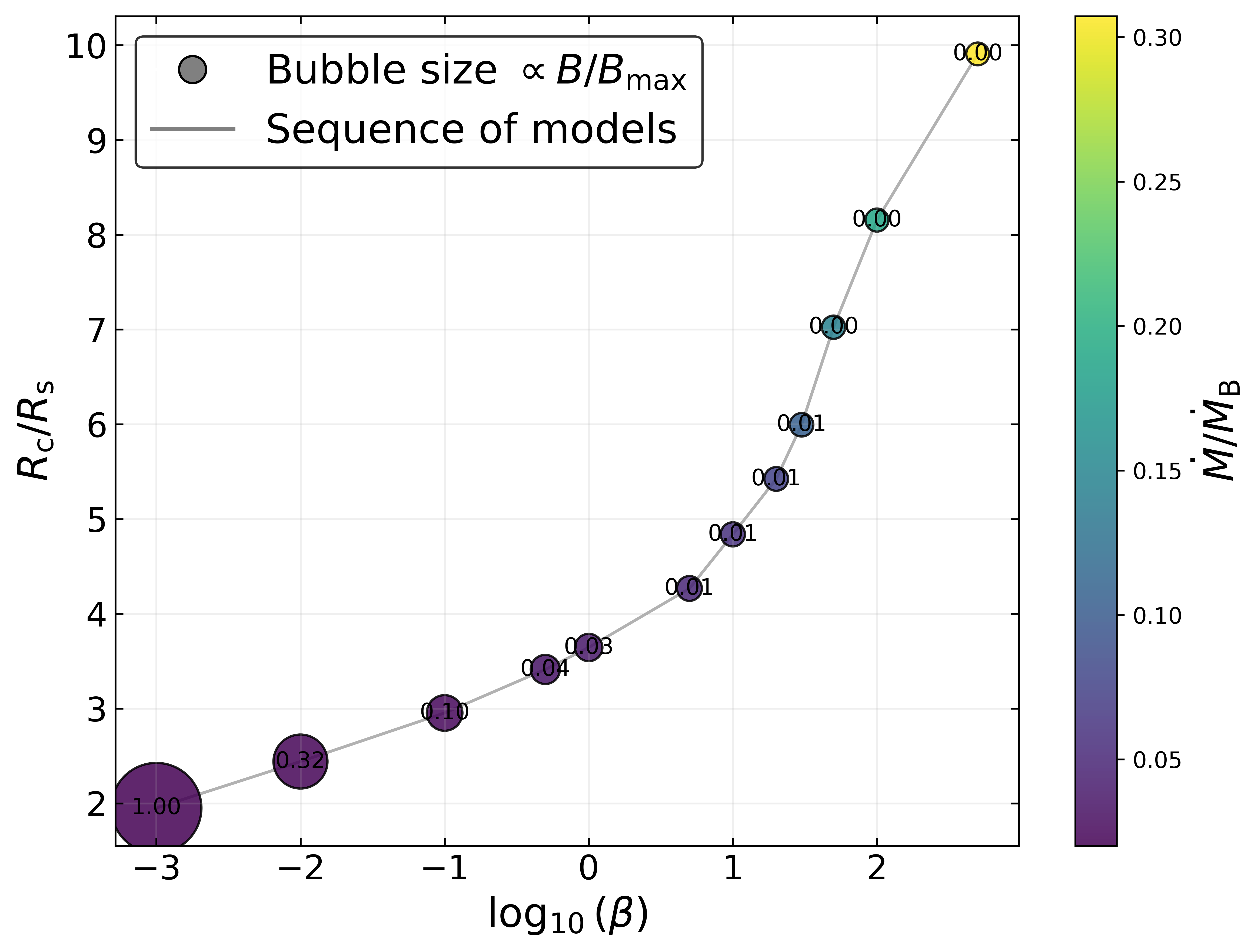}
    \hspace{0.01\textwidth}
    \includegraphics[width=0.4\textwidth]{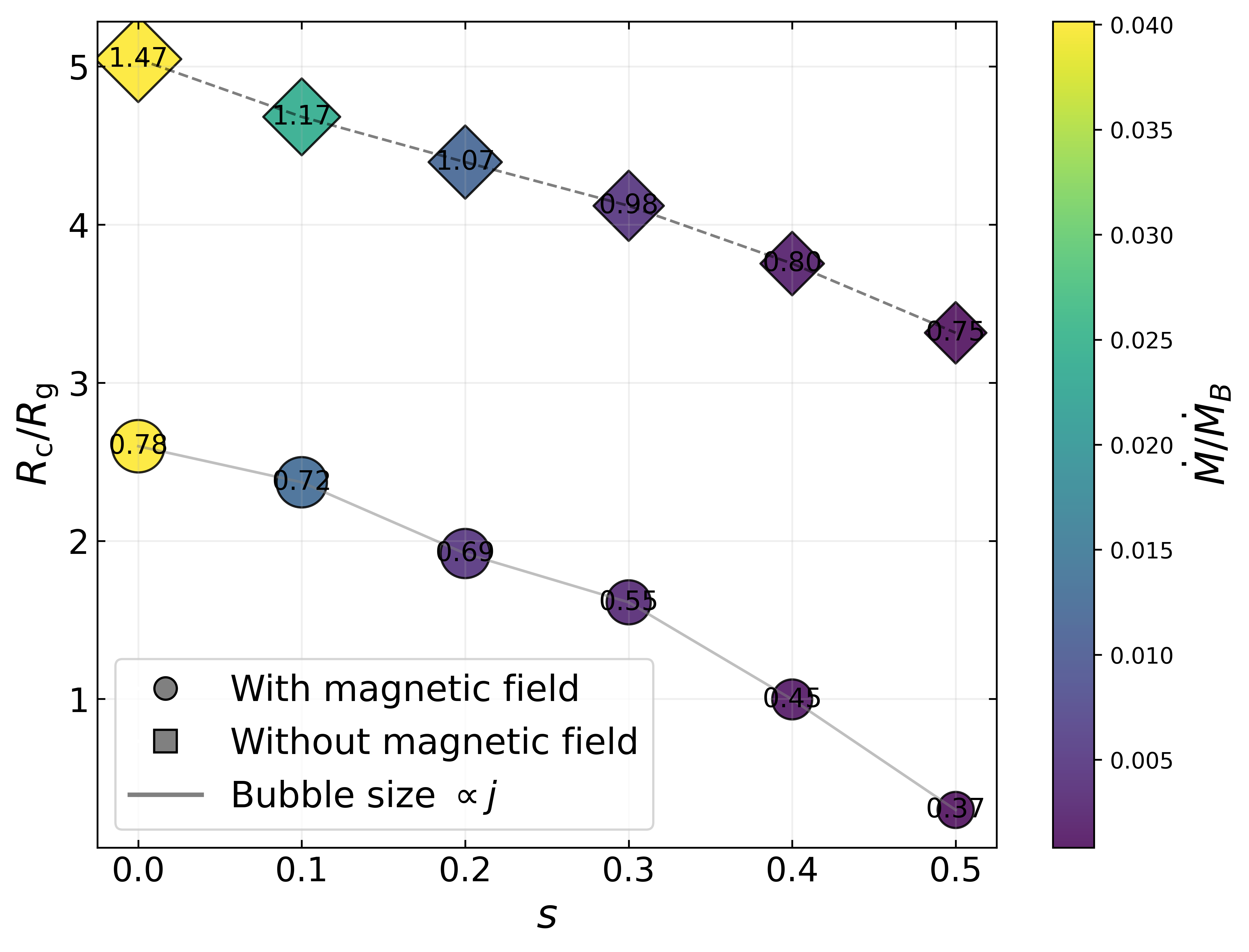}

    \caption{Left panel: Variation of the critical radius ($R_{\rm c}$) with the plasma beta parameter ($\beta=P_{\rm gas}/P_{\rm mag}$). The horizontal axis shows $\log_{10}(\beta)$, the bubble size is proportional to the normalized magnetic field strength ($B/B_{\rm max}$), and the color indicates the normalized mass accretion rate ($\dot{M}/\dot{M}_{\rm B}$). The numerical labels denote $B/B_{\rm max}$, and the gray line connects successive models. Here, $B_0=0.2$ and $s=0$ and $L = 85$. Right panel: Comparison of the critical radius ($R_{\rm c}$) as a function of the outflow parameter $s$ for magnetized (circles) and non-magnetized (diamonds) accretion solutions. The symbol size represents the specific angular momentum $\rm j$, and the color indicates the normalized mass accretion rate ($\dot{M}/\dot{M}_{\rm B}$). Magnetic fields reduce the critical radius and specific angular momentum, while increasing $\rm s$ decreases both the critical radius and the accretion rate.}
    \label{color2}
\end{figure*}

Figure \ref{Fig2} compares the radial structure of the Bondi-like and \ac{ADAF}-like accretion solutions in the absence and presence of magnetic fields. The upper-left panel shows the radial velocity profile. In both accretion branches, the magnetized solutions exhibit systematically lower inflow velocities than their hydrodynamic counterparts. The reduction is particularly evident in the Bondi-like solution, indicating that magnetic pressure provides additional support against gravitational infall and slows the radial motion of the gas.

The density profiles, shown in the upper-right panel, decrease approximately as power laws with increasing radius. The magnetized solutions remain less dense than the corresponding hydrodynamic models over the entire computational domain. The magnetized solutions remain less dense than the corresponding hydrodynamic models over the entire computational domain, consistent with the flattened density profile found by \citet{Li2025} in their independent flux-transport model of \ac{MAD} formation. This reduction is more pronounced in the Bondi-like branch, reflecting the lower mass accretion rate in the presence of magnetic pressure.

The lower-left panel presents the relative scale height, $ H/R$. Magnetic fields modify the vertical structure of the flow by reducing the geometrical thickness compared with the purely hydrodynamic solutions. Although both branches remain geometrically thick at small radii, the magnetized Bondi-like solution exhibits the largest decrease in $ H/R$, whereas the \ac{ADAF}-like solution is only moderately affected.

The radial temperature profiles are displayed in the lower-right panel. Both Bondi-like and \ac{ADAF}-like solutions become progressively cooler in the magnetized case, with the largest temperature differences occurring at intermediate radii. Since the thermal pressure is partially replaced by magnetic pressure, the gas requires a lower temperature to maintain force balance, leading to systematically reduced temperatures throughout the flow.

The sonic points, marked by filled circles, shift slightly in the magnetized solutions relative to the hydrodynamic cases, indicating that magnetic fields modify the transonic structure of the accretion flow. Overall, magnetic pressure produces slower, cooler, and less dense accretion flows while preserving the qualitative distinction between the Bondi-like and \ac{ADAF}-like branches.

Figure \ref{color2} (left panel) illustrates the dependence of the critical radius on the plasma beta parameter, $\beta=P_{\rm gas}/P_{\rm mag}$. As $\rm \beta$ increases, corresponding to a weaker magnetic field, the critical radius increases monotonically from $R_{\rm c}\simeq1.95\, R_{\rm s}$ to $R_{\rm c}\simeq 9.91\, R_{\rm s}$. At the same time, the normalized mass accretion rate, represented by the color scale, increases significantly from approximately $0.02$ to $0.31$. The bubble size and the numerical labels indicate the normalized magnetic field strength, $B/B_{\rm max}$, which decreases with increasing $ \beta$, consistent with the relation $B\propto \rm \beta^{-1/2}$. These results demonstrate that stronger magnetic fields (smaller $ \beta$) are associated with smaller critical radii and lower mass accretion rates, whereas weaker magnetic fields shift the critical point outward and enhance the accretion rate.

Figure \ref{color2} (right panel) compares the properties of accretion solutions in the presence and absence of magnetic fields. The critical radius $R_{\rm c}$
 decreases systematically with increasing outflow parameter $ s$ for both cases, indicating that stronger mass loss causes the sonic transition to occur closer to the central object. However, the magnetized solutions exhibit significantly smaller critical radii compared with the non-magnetized case. For example, at $ s=0$, the critical radius decreases from $R_{\rm c} \sim 5.05 \, R_{\rm s}$
 without magnetic fields to $R_{\rm c} \sim 2.60 \, R_{\rm s}$ when magnetic effects are included. This trend remains throughout the explored parameter range, suggesting that magnetic stresses enhance angular momentum transport and modify the dynamical structure of the accretion flow.
The bubble sizes indicate the variation of the specific angular momentum $j$. The non-magnetized solutions consistently have larger values of $j$ compared with the magnetized models, implying that magnetic fields efficiently remove angular momentum from the flow. Additionally, the color distribution shows that the normalized mass accretion rate $\dot{M}/\dot{M}_{\mathrm{B}}$
 decreases with increasing $ s$, as expected due to enhanced mass loss. At a given $ s$, the non-magnetized solutions generally maintain larger accretion rates, whereas magnetic fields reduce the accretion efficiency by altering the angular momentum transport and flow dynamics.

\begin{figure*}[htbp]
    \centering

    % First row
    \includegraphics[width=0.31\linewidth]{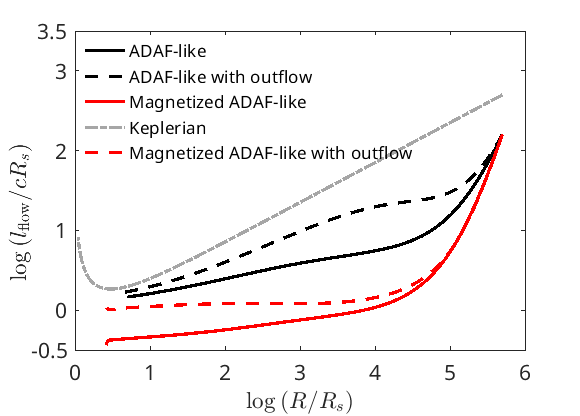}
    \hfill
    \includegraphics[width=0.3\linewidth]{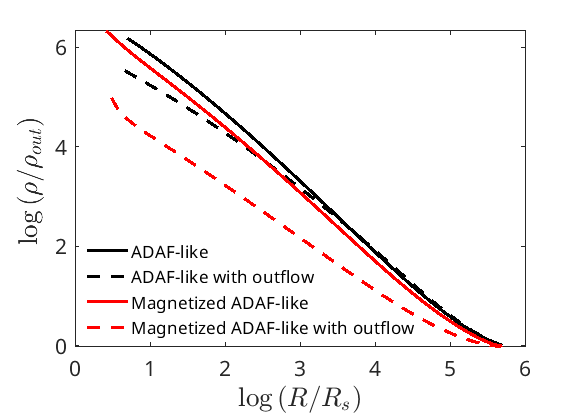}
    \hfill
    \includegraphics[width=0.3\linewidth]{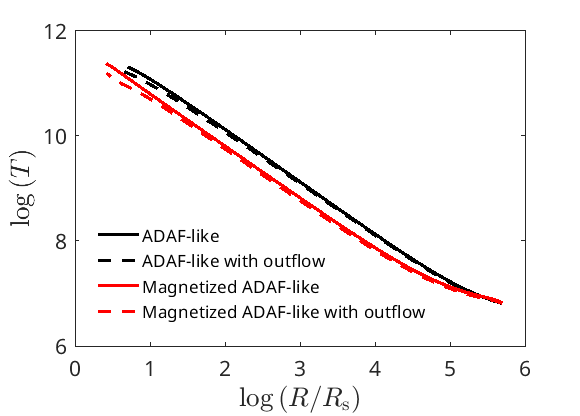}

    \vspace{0.1cm}

    % Second row
    \includegraphics[width=0.29\linewidth]{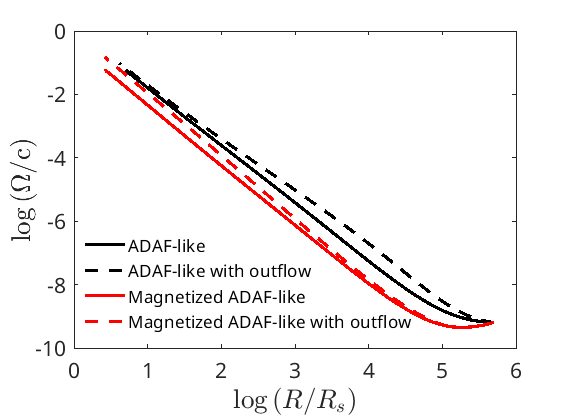}
    \hfill
    \includegraphics[width=0.3\linewidth]{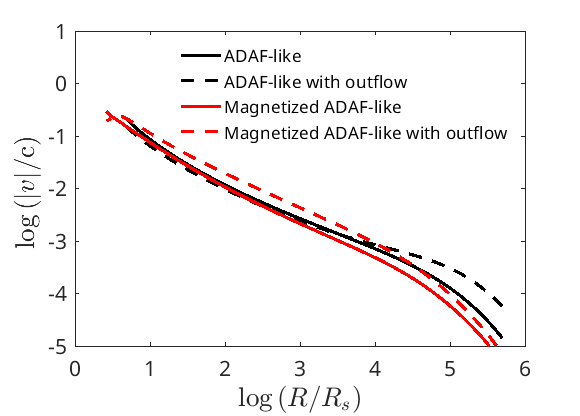}
    \hfill
    \includegraphics[width=0.3\linewidth]{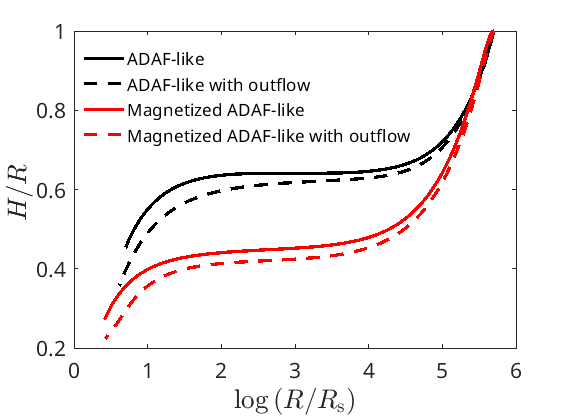}

    \caption{
Radial profiles of the physical quantities for accretion flows with an
outflow strength of $ s=0.3$ and plasma beta $ \beta=0.5$. The panels show
the specific angular momentum ($ l_{\rm flow} = \Omega R^2$), density ($ \rho$), temperature ($ T$), radial velocity ($ v$), and relative disk thickness
($ H/R$) as functions of radius. Black curves correspond to the \ac{ADAF}-like
solutions, while red curves represent the magnetized \ac{ADAF}-like solutions.
Solid and dashed lines denote models without and with outflow, respectively.
}
    \label{fig:six_plots}
\end{figure*}

Figure \ref{fig:six_plots} presents the radial structure of the accretion flow for an outflow strength of $\rm s=0.3$ and a moderately magnetized plasma ($\rm \beta=0.5$). Introducing magnetic fields significantly modifies the flow structure compared with the non-magnetized \ac{ADAF}-like solution.
Since the outflow is primarily important from $10\, {R\rm s}$ to the outer boundary, we restrict our analysis to this region. At smaller radii, the strong gravitational field of the \ac{BH} dominates the dynamics and suppresses the outflow near the horizon.

The magnetized models exhibit lower specific angular momentum over most of the radial range, although the profiles converge near the outer boundary. The density decreases monotonically with radius in all models, while the presence of both magnetic fields and outflow further reduces the gas density because mass is continuously removed from the accretion flow.

The temperature also decreases outward, with magnetized solutions remaining systematically cooler than the corresponding non-magnetized models. Consistent with this behavior, the sound speed is reduced in the magnetized flow, reflecting the lower thermal pressure.

The radial velocity gradually decreases with radius in all cases. The inclusion of magnetic fields results in slightly smaller inflow velocities, and the outflow further reduces the magnitude of the radial velocity, particularly at large radii.

The thickness of the disk ($\rm H/R$) increases toward the outer region in every model. However, magnetization produces a geometrically thinner disk than the purely hydrodynamic \ac{ADAF}-like solution. The addition of outflow slightly decreases the disk thickness throughout most of the computational domain. Overall, the combined effects of magnetic fields and outflow lead to a cooler, less dense, and geometrically thinner accretion flow while preserving the general radial trends of the \ac{ADAF}-like solution.

\subsection{2D simulations}\label{sec:3.1}

In this section, we present two-dimensional simulations of magnetized
accretion flows and compare our semi-analytical solutions in the
Newtonian framework with general-relativistic simulations performed
with the \textsc{BHAC} code \cite{Porth2017} for qualitative comparisons/numerical consistency checks. We consider two complementary setups:
magnetized Bondi accretion and accretion from a weakly magnetized
Fishbone--Moncrief torus around a non-rotating \ac{BH}. These
simulations allow us to test our semi-analytical solutions in a
relativistic setting and to investigate the effects of magnetic fields
on the accretion flow.

\subsubsection{Magnetized Bondi accretion}

The 2D Bondi simulation considers axisymmetric spherical accretion
onto a non-rotating \ac{BH}. We employ spherical modified Kerr-Schild (MKS) coordinates \citep{McKinney2004}, with $a=0$. In these coordinates, the grid is logarithmic in radius, so cells are small near the black hole and get bigger farther out, and in angle, more cells are placed near the equator. The resolution of
$\rm N_{\rm r} \times \rm N_{\rm \theta}=200\times100$, and a computational domain
extending from $ R=1.8 \, \rm M$ to $ R=10^5 \, \rm M$, with $\theta\in[0,\pi]$. The fluid obeys an ideal-gas equation of state
with an adiabatic index of $\gamma=5/3$. The sonic radius is located
at $ R_{\rm c}=8 \, \rm M$. The magnetic field is normalized such that the maximum
magnetization is $\sigma=  B^2/ \rho = 10^3$.

\begin{figure}
    \centering
    \includegraphics[width=1.0\linewidth]{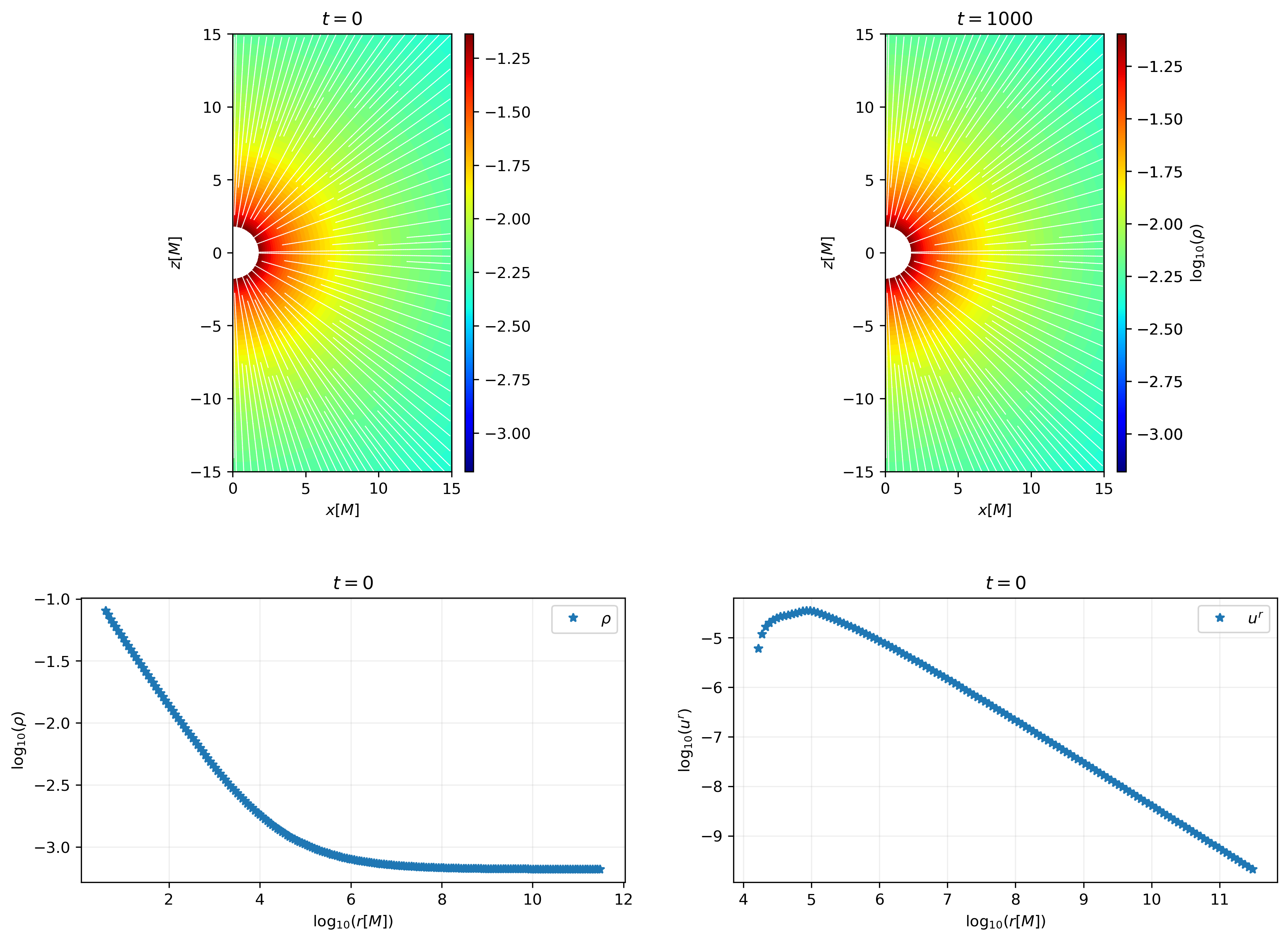}
    \caption{ Magnetized Bondi flow: Top: logarithmic density and streamlines in 2D magnetised Michel accretion at
times $t = 0$ M (left) and $t = 1000$ M (right). The horizon is marked by the black line at $R = 1.8 \, M$. Bottom: profiles at $\theta = \pi/2$ of,
from left to right, radial 3-velocity, density and radial magnetic field at $t = 0$ M (blue circles) and $t = 1000$ M.}
    \label{2Ds}
\end{figure}

Figure \ref{2Ds} shows the evolution of the magnetized
spherical accretion flow in the 2D simulations. The upper panels show
the logarithmic rest-mass density, $\log_{10}(\rho)$, together with the
magnetic-field streamlines at $ t=0$ (left) and $t=1000 \, \rm M$ (right). The
flow remains approximately spherical and the magnetic-field
configuration remains predominantly radial throughout the evolution.

The lower panels show the corresponding one-dimensional profiles of
the rest-mass density and radial four-velocity, $\rm u^r$, along the
equatorial plane, $ \theta=\pi/2$, at $ t=0$. The density decreases
outwards from the \ac{BH} and approaches an approximately constant
value at large radii, while the magnitude of the radial velocity
increases toward the \ac{BH}. The close similarity of the
two-dimensional density distributions at $ t=0$ and $ t=1000\, \rm M$ indicates
that the flow reaches and maintains a nearly stationary configuration
over the simulated time interval.

\subsubsection{Magnetized Fishbone--Moncrief torus}

To compare our semi-analytical solutions in a setting closer to our
intended applications, we also simulate accretion from a magnetized
and perturbed Fishbone--Moncrief torus around a non-rotating black
hole with $ M=1$ and $ a=0$. We consider a hydrodynamic equilibrium
torus threaded by a weak magnetic-field loop. The equilibrium torus
solution with constant angular momentum was originally presented by
Fishbone and Moncrief (\cite{Fishbone1976}) and Kozlowski et al.
(\cite{Kozlowski1978}), and has since become a standard initial
condition for GRMHD simulations.

Since our semi-analytical model is constructed for a non-rotating
\ac{BH}, we set the dimensionless spin parameter to $ a=0$,
corresponding to a Schwarzschild spacetime. We adopt an ideal-gas
equation of state with an adiabatic index of $\gamma=5/3$.

\begin{figure}
    \centering
    \includegraphics[width=1.0\linewidth]{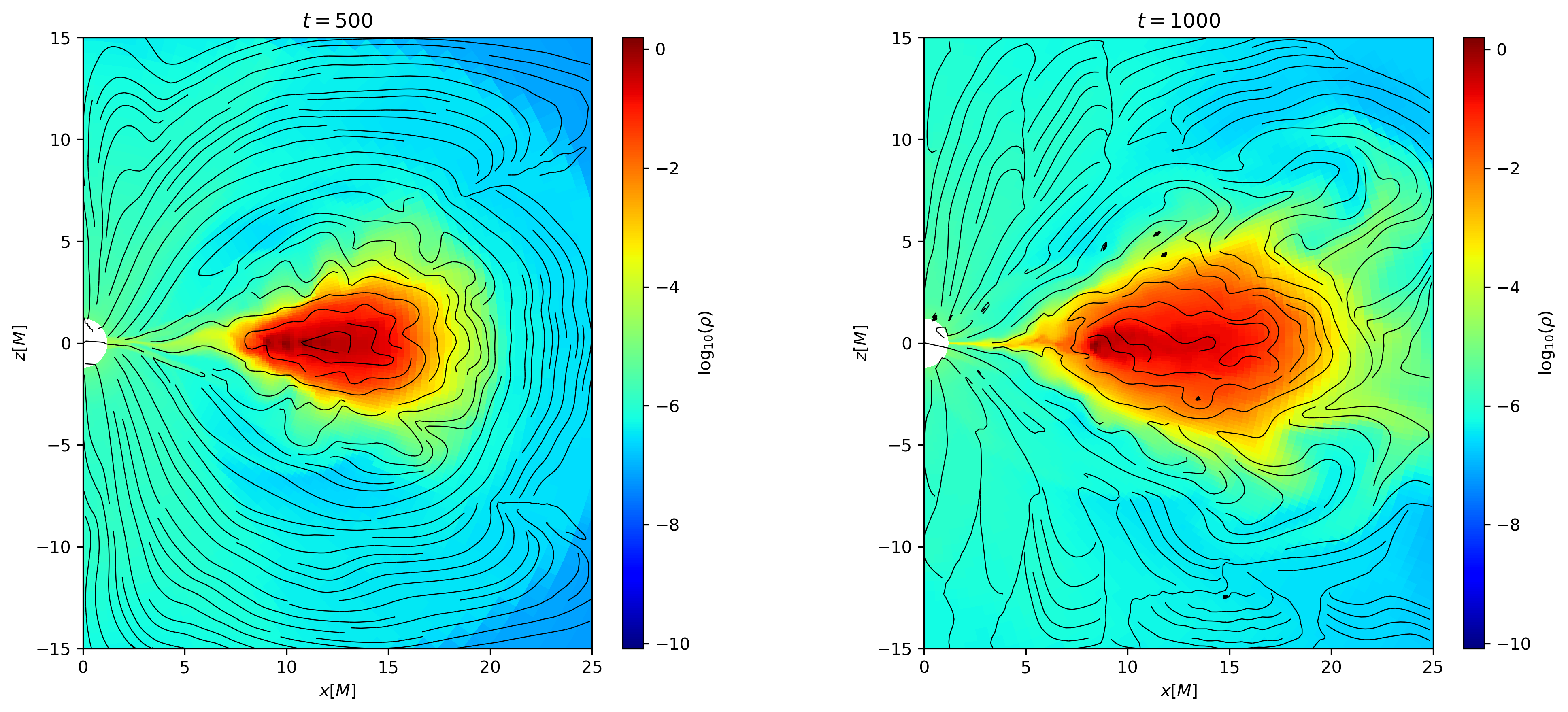}
     \includegraphics[width=0.8\linewidth]{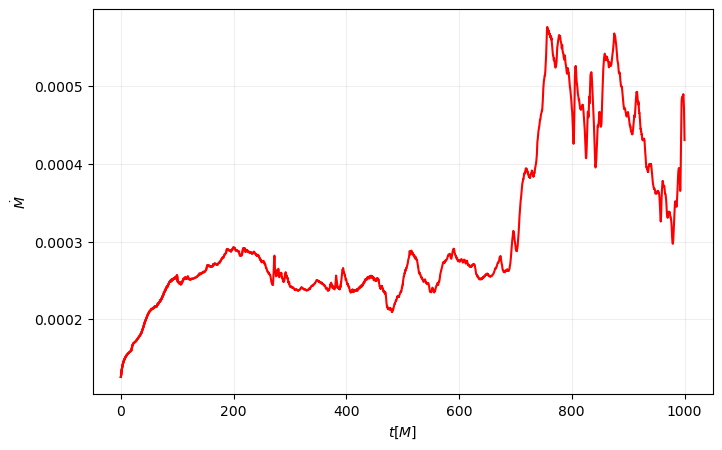}
    \caption{ Magnetised torus: Evolution of the 2D magnetised torus with resolution $192 \times 96$ for times $t/M \in \{500, 1000\}$ density and magnetic field lines $t = 500$ and $t=1000$ M (top). Accretion
rates in the 2D validation runs (bottom).}
    \label{simulation}
\end{figure}

Figure \ref{simulation} illustrates the qualitative time evolution
of the torus through the rest-mass density $\rho$. The computational
domain extends from $ r=1.2 \, \rm M$ to $ r=2500 \, \rm M$ and
$ \theta\in[0,\pi]$, with a resolution of
$\rm N_{\rm r}\times \rm N_{\rm \theta}=192\times96$. The system is evolved until
$ t=1000 \, \rm M$.

Following the initial transient, the magnetic field triggers the
magnetorotational instability (MRI), leading to the development of
turbulence and accretion onto the central \ac{BH}. The mass
accretion rate through the horizon subsequently approaches a
quasi-stationary state, as illustrated in Figure \ref{simulation}.

\subsection{Comparison with simulation}

We compare the semi-analytical predictions of Section~ \ref{sec:3} with the 2D
\textsc{BHAC} simulations of Section~ \ref{sec:3.1}. The semi-analytical Bondi-like
solutions place the critical radius in the range $ R_{\rm c} \simeq 400$--$470$ and disk-like are in range
$ R_{\rm c}\simeq 1.95$--$9.91\,R_{\rm s}$ depending on field strength
(Fig.~\ref{color2}, left panel); the 2D magnetized Bondi simulation,
evolved independently in a fully relativistic framework, converges to
$  R_{\rm c}=8 \, \rm M=4\, R_{\rm s}$, which falls within range of disk-like one. 

The 2D magnetized Bondi simulation converges to a sonic radius of $ R_{\rm c} = 8 \, \rm M = 4\,  R_{\rm s}$. This lies well below the ADAF-like semi-analytical range explored in Fig.~\ref{color2} $ R_{\rm c}\simeq 1.95$--$9.91\, R_{\rm s}$ and closer to, though smaller than, the Bondi-like range $R_{\rm c} \simeq 400$--$470$ shown in Fig.~1. Since the beta-scan in Fig.~\ref{color2} was carried out at $ L =85 $, a direct quantitative comparison at matched angular momentum is left for future work; at this stage the simulation confirms only that magnetic fields shift the sonic point inward relative to the unmagnetized case, in qualitative agreement with the semi-analytical trend.
The flattened
density profile obtained semi-analytically ($\rho\propto R^{-1}$ inside
the Bondi radius) is also consistent with the equatorial
density profile extracted from the 2D Bondi simulation.  Some differences are expected: the semi-analytical model is
Newtonian, steady, and height-integrated, while the \textsc{BHAC}
simulations are relativistic, time-dependent, and do not assume
height-integration; the semi-analytical model also does not follow the
time-dependent flux accumulation that drives \ac{MAD} formation, which
the torus simulation captures self-consistently. Despite these
differences, the agreement in critical radius and density structure
supports the semi-analytical prescription as a reasonable steady-state
description of the time-averaged flow.

Compared with the corresponding hydrodynamic solution, the magnetized
models also exhibit a shallower density profile and a substantially
lower mass accretion rate, consistent with previous numerical studies
\citep{Cho2023, Cho2023b}. The inner flow becomes strongly magnetized
and exhibits properties reminiscent of the \ac{MAD} regime
(\citealp{Igumenshchev2003, Narayan2003, Tchekhovskoy2011};
see also \cite{BisnovatyiKogan1974}). \citet{Ressler2020} similarly
found that the stellar wind-fed accretion flow in their model of
Sgr~A* approached the \ac{MAD} state. In our case, the flow remains
strongly magnetized over a large radial range, from the Bondi radius
($ R_{\rm B}\sim 5 \times 10^5\, R_{\rm s}$) to the vicinity of the event horizon
($2\, R_{\rm s}$). Unlike most GRMHD simulations in the literature, all
our simulations assume a non-spinning \ac{BH} ($ a = 0$), so the
spacetime has no preferred rotation axis to guide the infalling gas; a
spinning \ac{BH} could instead dictate the flow's rotational dynamics
through frame-dragging at small radii, a question we leave for future
work.

\section{Conclusions} \label{sec:4}

In this paper, we have obtained global solutions for Bondi-like and \ac{ADAF}-like accretion flows in the presence of magnetic fields, extending from the event horizon of the central supermassive black hole to galactic scales (the Bondi radius).

Adding a magnetic field of realistic strength to the external gas changes the results drastically. Inside the Bondi radius the density profile becomes $\rho\propto R^{-1}$ (instead of the classic Bondi/\ac{ADAF} slope $\rho\propto R^{-3/2}$), and the mass accretion rate onto the \ac{BH} drops to $\dot{M}\simeq 5\times10^{-3}\,\dot{M}_{\mathrm{B}}$ for case $\rm s = 0.5$.
This value is consistent with the accretion rates inferred for M87* ($\dot{M}\sim10^{-3}\,M_\odot\,{\rm yr}^{-1}$) versus ($\dot{M}_{\mathrm{B}}\sim 0.2\,M_\odot\,\mathrm{yr}^{-1}$); \citep{EHT2021, Russell2015} and for Sgr~A* $\sim10^{-4}$--$10^{-2}\,\dot{M}_{\mathrm{B}}$; \citep{Wang2013}. The same $ \rho\propto \rm R^{-1}$ scaling has been reported in several previous GRMHD simulations of rotating flows (\cite{Ressler2020, Chatterjee2022}) and even in some hydrodynamical calculations (\cite{Guo2023}), and is compatible with X-ray and EHT constraints on both sources.

Magnetic flux is efficiently transported inward through the \ac{ADAF} and accumulates in a region extending to several tens of Schwarzschild radii outside the horizon. In this region the flow is highly sub-Keplerian ($\Omega\sim(0.4$--$0.5)\,\Omega_{\rm K}$)
and magnetically dominated ($ \beta<1$), in agreement with earlier studies (\citealt{XieZdziarski2019, Chatterjee2022, Ressler2023}) and closely consistent with the independent semi-analytical flux-transport model of \citet{Li2025}, who found a \ac{MAD} region forming at a comparable radial extent with a nearly identical sub-Keplerian rotation rate.
We also find that stronger outflows (power-law index $s$ increasing from $0.1$ to $0.3$) enlarge the \ac{MAD} zone, suggesting that mass loss may play an important role in the formation and maintenance of a \ac{MAD}. The detailed contribution of magnetic outflows to \ac{MAD} assembly deserves further investigation.

When the accreting gas is magnetized, the black-hole magnetosphere saturates with a strong field. Energy is transported across these widely disparate scales by turbulent convection driven by magnetic reconnection near the \ac{BH}. Thus strong magnetic fields that accumulate on horizon scales transform the flow dynamics far from the SMBH and naturally explain both the extremely low observed accretion rates and at least part of the required energetic feedback.

Our present study is restricted to non-spinning \acp{BH} ($ a=0$). For a spinning \ac{BH}, frame dragging may significantly modify the rotational dynamics of the accreting gas, particularly in the inner region, and could therefore alter the magnetic-field structure and accretion properties. Investigating the effects of black-hole spin will be considered in future work.

An important additional feedback channel that remains to be explored quantitatively is the Blandford--Znajek mechanism (\cite{Blandford1977}), whereby magnetic fields threading a spinning \ac{BH} extract rotational energy and launch relativistic jets. In future work we will investigate the impact of black-hole spin and jet launching on the global accretion dynamics and feedback efficiency.

\section*{Acknowledgements}

I thank Sh. Abbassi and H. Olivares for their valuable comments, constructive suggestions, and fruitful discussions. In particular, I acknowledge H. Olivares for his helpful discussions and insights regarding the GRMHD simulations, which contributed to improving this work.

\section*{Data Availability}

No new data were generated or analysed in support of this research.



\bibliographystyle{mnras}
\bibliography{example} % if

\end{document}